\documentclass[smallcondensed]{svjour3}

\usepackage{amsmath}
\usepackage{natbib}
\usepackage{graphicx}
\usepackage{mathptmx}

\journalname{Celestial Mechanics and Dynamical Astronomy}

\newcommand{\order}[1]{\mathcal{O}\left(#1\right)}

\bibpunct{(}{)}{;}{a}{}{,}

\begin{document}

\title{Variational Integrators for Almost-Integrable Systems}
\author{Will M. Farr}

\institute{Kavli Institute for Astrophysics and Space Research, MIT
  Room 37-624C, 77 Massachusetts Ave., Cambridge, MA
  02139\\\email{farr@mit.edu}}

\maketitle

\abstract{We construct several variational integrators---integrators
  based on a discrete variational principle---for systems with
  Lagrangians of the form $L = L_A + \epsilon L_B$, with $\epsilon \ll
  1$, where $L_A$ describes an integrable system.  These integrators
  exploit that $\epsilon \ll 1$ to increase their accuracy by
  constructing discrete Lagrangians based on the assumption that the
  integrator trajectory is close to that of the integrable system.
  Several of the integrators we present are equivalent to well-known
  symplectic integrators for the equivalent perturbed Hamiltonian
  systems, but their construction and error analysis is significantly
  simpler in the variational framework.  One novel method we present,
  involving a weighted time-averaging of the perturbing terms, removes
  all errors from the integration at $\order{\epsilon}$.  This last
  method is implicit, and involves evaluating a potentially expensive
  time-integral, but for some systems and some error tolerances it can
  significantly outperform traditional simulation methods.


  }

\section{Introduction}

Symplectic integrators have been used since their introduction in
\citet{Wisdom1991} for simulations of gravitational systems which are
dominated by a central body.  These integrators split the Hamiltonian
for a system into two parts:
\begin{equation}
  H = H^{(A)} + \epsilon H^{(B)},
\end{equation}
where $H^{(A)}$ represents the influence of the dominant central body
on the bodies that orbit it, and $\epsilon H^{(B)}$, $\epsilon \ll 1$,
represents the mutual interactions of the bodies in orbit around it.
(In our solar system, $\epsilon \sim 10^{-3}$; for stars in orbit
around a central galactic black hole, $\epsilon \sim 10^{-6}$.)  These
integrators are a composition of evolutions under the separate pieces
of the Hamiltonian, which are individually integrable.  The
integrators in \citet{Wisdom1991} have a trajectory error which scales
as $\order{\epsilon h^3}$ over a single step of size $h$.
\citet{McLachlan1995}, \citet{Chambers2000}, and \citet{Laskar2001}
present improvements to the basic leapfrog scheme in
\citet{Wisdom1991} which involve more composition steps to eliminate
error terms at $\order{\epsilon}$; these integrators have errors after
a single step of size $h$ which scale as $\order{\epsilon^2 h^3} +
\order{\epsilon h^{n+1}}$ and are known as pseudo-high-order
integrators.  \citet{Wisdom1996} introduced correctors which can
completely eliminate errors at $\order{\epsilon}$ from the
integration.

In this work, we present integrators derived from a Lagrangian of the
form 
\begin{equation}
  L = L^{(A)} + \epsilon L^{(B)}
\end{equation}
based on a discrete variational principle which incorporates the
dominant $L^{(A)}$ motion of the system.  These variational
integrators subsume traditional symplectic integrators.
Section \ref{Sec:VariationalIntegrators} provdies a brief introduction
to variational integrators; \citet{Marsden2001} provides a thorough
mathematical grounding for these integrators and discusses in detail
their properties.  \citet{Lew2004} provides another introduction to
the topic, and discusses the use of variational integrators in a
space-time (PDE) context.  \citet{Lee2007} demonstrate a geometrically
exact method for simulating full-body dynamics in orbital mechanics
with a variational integrator.

In this paper, we derive the pseudo-high-order integrators presented
in \citet{McLachlan1995}, \citet{Chambers2000}, and \citet{Laskar2001}
using the variational framework.  The occurrence of the Gauss-Lobatto
quadrature coefficients in the composition formulas for these
integrators is a natural consequence in this framework of using
Gauss-Lobatto quadrature to approximate the contribution of $L^{(B)}$
to the action integral.  We exploit the flexibility of the variational
framework to derive a novel implicit integrator which eliminates all
errors from the trajectory at $\order{\epsilon}$ through averaging of
the perturbing Lagrangian, $\epsilon L^{(B)}$, along the trajectory of
$L^{(A)}$.  We present numerical evidence in Section
\ref{ExampleCalculationsSection} that this latter integrator can be
more efficient than standard symplectic integrators for some systems
of physical interest in spite of its higher cost per step.  We argue
in Section \ref{OrderEpsilon2Method} that this method should be more
stable at large stepsize for eccentric systems than the standard
symplectic integrators, even with symplectic correctors.

In this work we use the notation of \citet{Sussman2001}.  The function
which is the derivative of a function $f$ is denoted $Df$; if $f$
takes a vector argument $x$, then $Df(x)$ is a co-vector.  Similarly,
we denote the function which is the partial derivative with respect to
the $i$th argument (counting from zero) of a function, $g$, of
multiple arguments by $\partial_i g$.  Some examples relating our
notation to traditional notation:
\begin{eqnarray}
  Df(x) & = & \frac{df}{dx}(x) \textnormal{ or } \frac{\partial
    f}{\partial {\bf x}}({\bf x}) \\
  \partial_1 f(x,y,z) & = & \left. \frac{\partial f(x,b,z)}{\partial
      b}\right|_{b = y} \textnormal{ or } \left. \frac{\partial
      f({\bf x},{\bf b},{\bf z})}{\partial {\bf b}}\right|_{{\bf b} = {\bf y}}.
\end{eqnarray}
  
\section{Variational Integrators}
\label{Sec:VariationalIntegrators}

We can construct a variational integrator for a mechanical system with
Lagrangian $L(q,v)$---here assumed to be time-independent for
simplicity---by considering the action for the system over a small
interval of time \citep{Farr2007,Lew2004}:
\begin{equation}
  S\left(h,q_0, q_1\right) = \int_0^h dt\, L\left( q(t), Dq(t) \right),
\end{equation}
where $q(t)$ is the physical trajectory of the system with $q(0) =
q_0$ and $q(h) = q_1$.  We choose a function, called the
\emph{discrete Lagrangian}, which approximates the action integral:
\begin{equation}
  L_D\left(h, q_0, q_1 \right) \approx S\left(h, q_0, q_1\right).
\end{equation}
Then the \emph{discrete Euler-Lagrange equations}
\begin{eqnarray}
  \label{p0Eqn} \partial_1 L_D\left(h,q_0,q_1\right) & = & -p_0 \\
  \label{p1Eqn} \partial_2 L_D\left(h,q_0,q_1\right) & = & p_1
\end{eqnarray}
define an integrator, $\left(q_0,p_0\right) \mapsto
\left(q_1,p_1\right)$.  Depending on the structure of $L_D$, equation
\eqref{p0Eqn} may be implicit for $q_1$.  

For example, consider the Harmonic oscillator, which has Lagrangian 
\begin{equation}
  L = \frac{1}{2}\left( v^2 - q^2 \right)
\end{equation}
in suitable units.  Adopt the ansatz that $q(t) = q_0 + (q_1 - q_0)
(t-t_0)/h$; we can construct a discrete Lagrangian using the midpoint
rule:
\begin{equation}
  L_D\left(h,q_0,q_1\right) = h L\left( \frac{q_0+q_1}{2}, \frac{q_1 -
      q_0}{h} \right).
\end{equation}
The integrator equations \eqref{p0Eqn} and \eqref{p1Eqn} can be solved
explicitly for $q_1$ and $p_1$.  The result is
\begin{eqnarray}
  q_1 & = & \frac{4 - h^2}{4 + h^2} q_0 + \frac{4 h}{4+h^2} p_0 \\
  p_1 & = & -\frac{4 h}{4+h^2} q_0 + \frac{4-h^2}{4+h^2} p_0.
\end{eqnarray}
Comparing with the actual solution 
\begin{eqnarray}
  q(t) & = & q_0 \cos(t) + p_0 \sin(t) \\
  p(t) & = & - q_0 \sin(t) + p_0 \cos(t),
\end{eqnarray}
we see that the integrator follows the exact trajectory but with phase
error---that is, $q_1 = q(h + \delta t)$ and $p_1 = p(h + \delta t)$
for some phase error $\delta t$.

One way to understand variational integrators is to recall that the
action is an $F_1$-type generating function for the canonical
transformation that implements time-evolution (see, e.g.,
\citet[pp.~415--416]{Sussman2001}).  The action defines the $F_1$-type
map, $(q,p) \mapsto (Q,P)$, via
\begin{eqnarray}
  \partial_1 S\left(h,q,Q\right) & = & -p \\
  \partial_2 S\left(h,q,Q\right) & = & P;
\end{eqnarray}
these are just equations \eqref{p0Eqn} and \eqref{p1Eqn}, with the
discrete Lagrangian---an approximate action---replaced by the true
action.

Alternately, consider our approximation to the action over a longer
interval:
\begin{multline}
  S_{\textnormal{tot}}\left(q_0, q_1, q_2, \ldots \right) = S\left(h,
    q_0,q_1\right) + S\left(h, q_1, q_2 \right) + \ldots \\ \approx
  L_D\left(h, q_0, q_1 \right) + L_D\left(h, q_1, q_2 \right) +
  \ldots.
\end{multline}
Then equations \eqref{p0Eqn} and \eqref{p1Eqn} express the stationarity
of our approximation to the action with respect to the intermediate
positions $q_1$, $q_2$, ....
\begin{equation}
  \partial_1 S_{\textnormal{tot}}\left(q_0, q_1, q_2, \ldots \right)
  \approx \partial_2 L_D\left(h,q_0,q_1\right) + \partial_1
  L_D\left(h, q_1, q_2 \right) = p_1 - p_1 = 0.
\end{equation}

Because the mapping in equations \eqref{p0Eqn} and \eqref{p1Eqn} is
the extremization of an approximation to the action, it shares many of
the desirable properties of the exact trajectory which extremizes the
true action.  For example, if $L_D$ has symmetries, a discrete
N\"{o}ther's theorem implies that the corresponding momenta are
conserved.  Also, the mapping is symplectic: denote the mapping by $F$.
Then the symplectic form on phase space $dq \wedge dp$ is invariant
under pushforward by $F$:
\begin{equation}
F^* \left( dq \wedge dp \right) = dq \wedge dp.
\end{equation}
Finally, it is possible to show via backward error analysis
\citep{Lew2004} that the mapping implements the \emph{exact} evolution
for some Lagrangian $\bar{L}$ which is close to $L$.  Therefore, the
evolution under $L_D$ remains on a constant-energy surface in phase
space for $\bar{L}$.  Because $\bar{L}$ is close to $L$, the evolution
under $L_D$ always remains close to the constant-energy surface of
$L$, so the \emph{long-term} energy error of the mapping under $L_D$
is bounded.

The order of error in the mapping of equations \eqref{p0Eqn} and
\eqref{p1Eqn} is the same as the order of error in the action
approximation $L_D$ \citep[Theorem 2.3.1]{Marsden2001}.  That is, if
\begin{equation}
  L_D(h,q(0),q(h)) = S(h,q(0),q(h)) + \order{h^{n+1}},
\end{equation}
where $q(t)$ is a stationary-action trajectory, then the mapping
defined by equations \eqref{p0Eqn} and \eqref{p1Eqn} approximates the
physical trajectory to $\order{h^{n+1}}$, and therefore defines an
$n$-th order integrator for $L$.  We shall exploit this result in the
error analysis of the integrators presented in this paper.  

\subsection{Multi-Point Variational Integrators}

It is often advantageous to allow the discrete trajectory to pass
through intermediate points between $q_0$ and $q_1$.  For example, we
may imagine that the discrete trajectory is a polynomial in time which
interpolates between the positions $q_0$, $q_0'$, $q_0''$, ...,
$q^{(n)}_0$, $q_1$.  With intermediate positions, the discrete
Euler-Lagrange equations become
\begin{eqnarray}
  \partial_1 L_D\left(h,q_0, q'_0, q''_0, \ldots, q^{(n)}_0, q_1 \right) & = & -
  p_0 \\
  \partial_{i+1} L_D\left(h,q_0, q'_0, q''_0, \ldots, q^{(n)}_0, q_1 \right) & =
  & 0, \qquad i = 1,2,\ldots,n \\
  \partial_{n+2} L_D\left(h,q_0, q'_0, q''_0, \ldots, q^{(n)}_0, q_1
  \right) & = & p_1.
\end{eqnarray}
The intermediate equations express that the action is stationary with
respect to the intermediate positions, while the other equations give
the time-evolution canonical transformation.

We can always (in principle) re-express any multi-point discrete
Lagrangian as an equivalent two-point discrete Lagrangian by solving
the intermediate equations for the intermediate positions in terms of
the start and end positions:
\begin{equation}
  \tilde{L}_D\left(h, q_0, q_1 \right) = L_D\left(h, q_0,
    q'_0\left(h,q_0,q_1\right), \ldots,
    q^{(n)}_0\left(h,q_0,q_1\right), q_1 \right),
\end{equation}
where the functions $q'_0\left(h,q_0,q_1\right)$, ...,
$q^{(n)}_0\left(h, q_0,q_1\right)$ solve
\begin{equation}
  \partial_{i+1} L_D\left(h,q_0, q'_0, q''_0, \ldots, q^{(n)}_0, q_1
  \right) = 0, \qquad i = 1,\ldots,n.
\end{equation}
Applying equations \eqref{p0Eqn} and \eqref{p1Eqn} to $\tilde{L}_D$
gives the same trajectory as the discrete Euler-Lagrange equations for
$L_D$.

One way to form multi-point discrete Lagrangians is to chain together
two separate discrete Lagrangians:
\begin{equation}
  L_D\left(h,q_0,q_0',q_1\right) = L_D^{(1)}\left(h,q_0,q_0'\right) + L_D^{(2)}\left(h,q_0',q_1\right).
\end{equation}
A quick calculation shows that the evolution under $L_D$ is a
composition of evolution under $L_D^{(2)}$ with that under
$L_D^{(1)}$.  

A common way to construct variational integrators of various orders is
to assume that the trajectory of the system is a polynomial in time
which passes through some discretization points: $q = q\left(t; q_0,
  q_0', \ldots, q_0^{(n)}, q_1 \right)$.  One then forms a discrete
Lagrangian using a quadrature rule on the action integral evaluated on
the discrete trajectory:
\begin{multline}
  L_D\left(h, q_0, q_0', \ldots, q_0^{(n)}, q_1\right) \\ = h \sum_i
  w_i L\left(q\left(t_i; q_0, q_0', \ldots, q_0^{(n)}, q_1
    \right), \partial_0 q\left(t_i; q_0, q_0', \ldots, q_0^{(n)}, q_1
    \right) \right),
\end{multline}
where $\left\{w_i, t_i\right\}$ are the weights and times of a
quadrature rule.  The order of the resulting map is determined by the
orders of the polynomial interpolation and quadrature rule.  

\section{Almost-Integrable Systems}
\label{IntegratorSection}
An \emph{almost-integrable system} has a Lagrangian of the form
\begin{equation}
  L = L^{(A)} + \epsilon L^{(B)},
\end{equation}
where $\epsilon \ll 1$, and the trajectories of $L^{(A)}$ are
calculable analytically (or at least efficiently).  In contrast to the
general Lagrangian, where little can be said about trajectories, we
expect that the trajectories of $L$ are going to be, in some sense,
close to those of $L^{(A)}$.  We should take advantage of this fact
when designing variational maps to approximate the trajectories of $L$
instead of blindly assuming polynomial-in-time motion as discussed at
the end of the last section.

The integrators we are about to describe all use as a component a
particular discrete Lagrangian:
\begin{equation}
  L_D^{(A)}\left(h,q_A(0),q_A(h)\right) \equiv \int_0^h dt\, L^{(A)}\left(q_A(t),Dq_A(t)\right),
\end{equation}
where $q_A(t)$ is the trajectory corresponding to the Lagrangian
$L^{(A)}$.  $L_D^{(A)}$ is the exact action for the $A$-subsystem on
its trajectories; applying equations \eqref{p0Eqn} and \eqref{p1Eqn}
to $L_D^{(A)}$ gives the $q_A$ trajectory:
\begin{eqnarray}
  \partial_1 L_D^{(A)}\left(h,q_A(0),q_A(h)\right) & = & -p_A(0) \label{AIntegratorp0Eqn}\\
  \partial_2 L_D^{(A)}\left(h,q_A(0),q_A(h)\right) & = & p_A(h) \label{AIntegratorp1Eqn}.
\end{eqnarray}
In general, it is not necessary to compute $L_D^{(A)}$; all that is
necessary is to be able to efficiently compute the mapping defined by
equations \eqref{AIntegratorp0Eqn} and \eqref{AIntegratorp1Eqn}.

\subsection{Composition Maps Using Quadrature Rules}

In this subsection we will show how some well-known symplectic maps
are equivalent to variational integrators and we will analyze their
errors in the variational framework.  Both the derivation and error
analysis are considerably simpler in the variational framework.  We
will assume that
\begin{equation}
  L^{(B)}(q,v) = -V^{(B)}(q);
\end{equation}
this common case is required for the integrators we discuss to be
compositional.  

We will see that the well-known symplectic integrators in
\citet{McLachlan1995}, \citet{Chambers2000}, and \citet{Laskar2001}
result from using Gauss-Lobatto quadrature to approximate
\begin{equation}
  -\int_0^h dt\, V^{(B)}(q(t))
\end{equation}
in the discrete Lagrangian, assuming $q_A$ trajectories between the
quadrature points.  In \citet{McLachlan1995}, \citet{Chambers2000},
and \citet{Laskar2001} the coefficients for these mapping integrators
come from the solution to algebraic equations involving iterated
commutators of the Hamiltonians $H^{(A)}$ and $H^{(B)}$, and turn out
to be Gauss-Lobatto quadrature weights and times.  In the framework of this
paper, the Gauss-Lobatto quadrature coefficients arise naturally from the
attempt to approximate the time-integral of $V^{(B)}$ to a given
order.

\subsubsection{Kick-Drift-Kick Leapfrog}

Suppose we choose
\begin{eqnarray}
  L_D\left(h, q_0, q_1\right) & = & L_D^{(A)}\left(h, q_0, q_1 \right) +
  \epsilon L_D^{(B)}\left(h,q_0,q_1\right) \nonumber \\ 
  & = & L_D^{(A)}\left(h, q_0,
    q_1 \right) - \epsilon \frac{h}{2} \left[ V^{(B)}\left( q_0 \right) + V^{(B)}\left( q_1
    \right) \right].
\end{eqnarray}
Here we have simply used the trapezoidal rule (a two-point Gauss-Lobatto
quadrature)
\begin{equation}
  \int_a^b dx\, f(x) \approx \frac{b-a}{2} \left[ f(a) + f(b)\right]
\end{equation}
to approximate the contribution to the action from $L_B$.  Applying
equations \eqref{p0Eqn} and \eqref{p1Eqn} to this discrete Lagrangian,
we have
\begin{eqnarray}
  -p_0 & = & \partial_1 L_D^{(A)}\left(h, q_0, q_1 \right) - \epsilon
  \frac{h}{2} DV^{(B)}\left(q_0\right) \\
  p_1 & = & \partial_2 L_D^{(A)}\left(h, q_0, q_1 \right) - \epsilon
  \frac{h}{2} DV^{(B)}\left(q_1\right).
\end{eqnarray}
These can be re-written in a suggestive way:
\begin{eqnarray}
  -\left(p_0 - \epsilon \frac{h}{2} DV^{(B)}\left(q_0\right) \right) &
  = & \partial_1 L_D^{(A)}\left(h, q_0, q_1 \right) \\
  p_1 & = & \partial_2 L_D^{(A)}\left(h, q_0, q_1 \right) - \epsilon
  \frac{h}{2} DV^{(B)}\left(q_1\right).
\end{eqnarray}
The solution to the first equation is the $q_1$ that results from
evolving $\left(q_0, p_0 - \epsilon h/2
  DV^{(B)}\left(q_0\right)\right)$ forward by $L^{(A)}$.  The final
momentum is then $p_A - \epsilon \frac{h}{2}
DV^{(B)}\left(q_1\right)$.  In other words, we kick by $-\epsilon
\frac{h}{2} DV^{(B)}\left(q_0\right)$, drift by $L^{(A)}$, and then
kick by $-\epsilon \frac{h}{2} DV^{(B)}\left(q_1\right)$.  The
algorithm is kick-drift-kick leapfrog---see, e.g.\ \citet{Wisdom1991}.

Using the variational framework to analyze the error of
kick-drift-kick leapfrog, we consider 
\begin{equation}
  \Delta \equiv L_D\left(h, q(0), q(h)\right) - S(h, q(0), q(h)).
\end{equation}
The trajectory error of kick-drift-kick leapfrog will be of the same
order as $\Delta$.  Expanding, we have
\begin{multline}
  \label{KDKLDError}
  \Delta = \Delta_A + \epsilon \Delta_B \\ = \int_0^h dt\, \left[
    L_A\left( q_A(t; q(0), q(h)), Dq_A(t; q(0), q(h))\right) -
    L_A\left( q(t), Dq(t) \right) \right] \\ - \epsilon \left[
    \frac{h}{2} \left( V^{(B)}(q(0)) + V^{(B)}(q(h))\right) - \int_0^h
    dt\, V^{(B)}(q(t)) \right].
\end{multline}

For the first term, we have 
\begin{multline}
  \Delta_A = \frac{\delta}{\delta q_A} \left[ \int_0^h dt\, L_A\left(
      q_A(t; q(0), q(h)), Dq_A(t; q(0), q(h))\right) \right] \delta
  q_A \\ + \frac{\delta^2}{\delta q_A^2} \left[ \int_0^h dt\,
    L_A\left( q_A(t; q(0), q(h)), Dq_A(t; q(0), q(h))\right) \right]
  \delta q_A^2 + \order{\delta q_A^3},
\end{multline}
where $\delta q_A$ is the trajectory difference between $q_A$ and $q$.
$\delta q_A$ is $\order{\epsilon h}$ because the Lagrangian depends on
the velocity, and the two trajectories must differ at first order in
$h$ in their velocities (since they feel different forces at order
$\epsilon$).  But, the trajectory $q_A$ is the solution to the
Euler-Lagrange equations for $L_A$, so the first order variation of
$L_A$ vanishes on $q_A$, and only the second-order term contributes.
Therefore, we have
\begin{equation}
  \Delta_A = \order{\epsilon^2 h^3}.
\end{equation}

For the second term of equation \eqref{KDKLDError}, we have
\begin{equation}
  \epsilon \Delta_B = \epsilon \left[ \order{h^3} \right] =
  \order{\epsilon h^3},
\end{equation}
arising from the truncation error in the quadrature rule.  Putting the
two terms together, we see that 
\begin{equation}
  \Delta = \Delta_A + \epsilon \Delta_B = \order{\epsilon h^3} +
  \order{\epsilon^2 h^3},
\end{equation}
with the term at $\order{\epsilon}$ arising from the quadrature rule
error, and the term at $\order{\epsilon^2}$ arising from the error in
$L_D^{(A)}$.  This will be a general feature of the integrators in
this section: the quadrature error determines the $\order{\epsilon}$
integrator error, while the $\order{\epsilon^2}$ error is determined
by the error in the $L_D^{(A)}$ term.

\subsubsection{S4B}

Consider choosing a higher-order quadrature rule for the $\epsilon
L^{(B)}$ term.  For example, a three-point Gauss-Lobatto quadrature rule,
with $q_A$ trajectories between the quadrature points:
\begin{multline}
  L_D\left(h, q_0, q_0', q_1\right) = \\ L_D^{(A)}\left(\frac{h}{2},
    q_0, q_0'\right) + L_D^{(A)}\left(\frac{h}{2}, q_0', q_1 \right) -
  \epsilon \frac{h}{6} \left[ V^{(B)}\left(q_0\right) + 4
    V^{(B)}\left(q_0'\right) + V^{(B)}\left(q_1\right) \right]
\end{multline}
As above, this integrator can be written as a sequence of kicks and
drifts: a kick by $-\epsilon \frac{h}{6} DV\left(q_0\right)$, a drift
by $h/2$ with respect to $L^{(A)}$ to $q_0'$, a kick by $-\epsilon
\frac{2h}{3} DV\left(q_0'\right)$, a drift by $h/2$ with respect to
$L^{(A)}$ to $q_1$, and a final kick by $-\epsilon \frac{h}{6}
DV\left(q_1\right)$.

As before the error introduced by the $L_D^{(A)}$ part is second order
in $\delta q_A$, or $\order{\epsilon^2 h^3}$, while the quadrature
rule introduces an error in the $V^{(B)}$ part of $\order{\epsilon
  h^5}$.  (There is an additional error in the quadrature arising from
the trajectory error $\delta q_A$, which contributes at
$\order{\epsilon^2 h^3}$ because the potential is independent of the
velocity difference between $q$ and $q_A$.)

Thus, the error in this method scales as 
\begin{equation}
  \Delta = \order{\epsilon h^5} + \order{\epsilon^2 h^3}.
\end{equation}
It belongs to a class of integrators known as ``pseudo-high-order'',
first discovered by \citet{McLachlan1995}, and introduced to the
astronomical community in \citet{Chambers2000} and \citet{Laskar2001}.
These integrators are useful because, for small enough $\epsilon$,
they behave as high-order integrators, even though they are formally
second order.  The name of this section (and the integrator) is S4B,
from \citet{Chambers2000}.

\citet{McLachlan1995} originally derived the coefficients of the
drifts and kicks for this integrator by attempting to eliminate
commutator terms in the Lie series for Hamiltonian evolution; he noted
that the coefficients which eliminate the desired
first-order-in-$\epsilon$ terms are identical to the Gauss-Lobatto
quadrature coefficients.  In this work, we can understand this as a
consequence of attempting high-order quadrature of the contribution of
$L^{(B)}$ to the action.

\subsubsection{S6B}

Consider now the sixth-order Gauss-Lobatto quadrature for the
$L^{(B)}$ terms, interspersed with $q_A$ evolution:
\begin{multline}
  L_D\left(h, q_0, q_0', q_0'', q_1 \right) = \\
  L_D^{(A)}\left(\frac{h\left(5-\sqrt{5}\right)}{10}, q_0,q_0'\right)
  + L_D^{(A)}\left(\frac{h}{\sqrt{5}}, q_0', q_0'' \right) +
  L_D^{(A)}\left(\frac{h \left(5 - \sqrt{5}\right)}{10}, q_0'',
    q_1\right) \\ - \epsilon \frac{h}{12} \left[
    V^{(B)}\left(q_0\right) + 5 V^{(B)}\left(q_0'\right) + 5
    V^{(B)}\left(q_0''\right) + V^{(B)}\left(q_1\right) \right]
\end{multline}
This is exactly the sequence of drifts and kicks for the S6B
integrator \citep{Chambers2000}.  Once again, the error is a
combination of errors from the quadrature at $\order{\epsilon}$, and
errors from $L_D^{(A)}$ at $\order{\epsilon^2}$:
\begin{equation}
  \Delta = \order{\epsilon h^7} + \order{\epsilon^2 h^3}
\end{equation}

\subsection{An $\order{\epsilon^2}$ Method}
\label{OrderEpsilon2Method}

We can eliminate all errors at $\order{\epsilon}$ by using for
$L_D^{(B)}$ the exact integral of $V^{(B)}$ along the $q_A$
trajectory.  Define
\begin{multline}
  L_D\left(h, q_0, q_1 \right) = L_D^{(A)}\left(h, q_0, q_1\right) +
  \epsilon L_D^{(B)}\left(h, q_0, q_1\right) \\ = L_D^{(A)}\left(h,
    q_0, q_1\right) - \epsilon \int_0^h dt\, V^{(B)}\left(q_A\left(t;
      q_0, q_1\right)\right).
\end{multline}
Applying equation \eqref{p0Eqn} to $L_D$, and moving the $L_D^{(B)}$
term to the left-hand-side, we obtain
\begin{equation}
  -\left( p_0 + \epsilon \partial_1 L_D^{(B)}\left(h, q_0, q_1\right)
  \right) = L_D^{(A)}\left(h, q_0, q_1 \right),
\end{equation}
which we must solve for $q_1$.  The momentum kick,
\begin{equation}
  \epsilon \partial_1 L_D^{(B)}\left(h, q_0, q_1\right) = -\epsilon
  \int_0^h dt\, DV^{(B)}\left( q_A\left(t; q_0, q_1 \right)
  \right) \partial_1 q_A\left(t; q_0, q_1 \right),
\end{equation}
is the time-averaged force along an $L^{(A)}$ trajectory weighted by
$\partial_1 q_A\left(t; q_0, q_1 \right)$---in general, this weight
favors the initial periods of the trajectory, since $q_A\left(h; q_0,
  q_1\right) = q_1$ independent of $q_0$.  Once the point $q_1$ is
determined, the new momentum is
\begin{equation}
  p_1 = \partial_2 L_D\left(h, q_0, q_1 \right) = p_1^{(A)} +
  \epsilon \partial_2 L_D^{(B)}\left(h, q_0, q_1 \right).
\end{equation}
Here, the kick 
\begin{equation}
  \epsilon \partial_2 L_D^{(B)}\left(h, q_0, q_1 \right) = - \epsilon
  \int_0^h dt\, DV^{(B)}\left( q_A\left(t; q_0, q_1 \right)
  \right) \partial_2 q_A\left(t; q_0, q_1 \right),
\end{equation}
is the time-averaged force along an $L^{(A)}$ trajectory weighted by
$\partial_2 q_A\left(t; q_0, q_1 \right)$---which tends to favor later
points in the trajectory, since $q_A\left(0; q_0, q_1\right) = q_0$
independent of $q_1$.  Because the momentum kicks are related to the
time-averaged force along the integrator trajectory, the integrator
has a flavor of averaging.  Timesteps with this integrator can be as
large as the time it takes the trajectory to deviate \emph{on average}
from $q_A$, in contrast to the integrators from the previous
subsection.  In those integrators timesteps must be small enough that
both $q_A$ adequately approximates the trajectory \emph{and} that the
sequence of kicks adequately approximates the averaged force.
Evaluating the averaged force in the way that the $\order{\epsilon^2}$
variational method does removes this second restriction.  This could
be important in the simulation of highly eccentric systems.

\subsubsection{Error Analysis}

The error from the $V^{(B)}$ integral is 
\begin{equation}
  \label{secondOrderBError}
  \epsilon \Delta_B = \epsilon \frac{\delta}{\delta q_A} \left[
    \int_0^h dt\, V^{(B)}\left(q_A\left(t; q_0, q_1\right)\right)
  \right] \delta q_A + \order{\delta q_A^2} = \order{\epsilon^2 h^3}.
\end{equation}
Because $V^{(B)}$ depends only on $q$ and not on $\dot{q}$, $\delta
q_A$ scales as $\order{\epsilon h^2}$ (the true trajectory $q$ and
$q_A$ must differ at order $\epsilon h^2$ because they feel different
forces of size $\epsilon$).

Combining the error in equation \eqref{secondOrderBError} with
$\Delta_A$, we obtain 
\begin{equation}
  \Delta = \order{\epsilon^2 h^3},
\end{equation}
resulting in a method which is formally second-order, but has no
errors at $\order{\epsilon}$.  

The method is implicit; one must solve equation \eqref{p0Eqn},
\begin{equation}
  -p_0 = \partial_1 L_D\left(h, q_0, q_1\right),
\end{equation}
for $q_1$.  This can be accomplished through Newton iteration, or
through the iterative method which follows.  (In practice, for small
$\epsilon < 10^{-3}$ and systems of modest dimension, we find
that the iterative method is more efficient than Newton iteration.)
Define the sequence $\left\{ q_1^{(i)} \right\}$ by 
\begin{equation}
  -\left( p_0 - \epsilon \partial_1 L_D^{(B)}\left(h, q_0, q_1^{(i-1)}
    \right) \right) = \partial_1 L_D^{(A)}\left(h, q_0, q_1^{(i)}\right).
\end{equation}
If $q_1^{(i-1)}$ is known, then $q_1^{(i)}$ is just the evolution of
the state $\left(q_0, p_0 - \epsilon \partial_1 L_D^{(B)}\left(h, q_0,
    q_1^{(i-1)} \right)\right)$ by $L_A$.  For small $\epsilon$, this
sequence $\left\{q_1^{(i)}\right\}$ converges linearly to the desired
$q_1$.

The efficiency of this method will depend on how tractable it is to
evaluate 
\begin{equation}
  \int_0^h dt\, V^{(B)}\left(q_A\left(t; q_0, q_1\right)\right)
\end{equation}
as a function of the endpoints $q_0$ and $q_1$.  In the next section,
we will examine the performance of the methods introduced in this
section on some example problems.

\section{Example Calculations}
\label{ExampleCalculationsSection}

In this section, we apply the integrators from the previous section to
some example problems.  

\subsection{The Perturbed SHO}

Consider the Lagrangian 
\begin{equation}
  L(q, v) = \frac{1}{2} \left( v^2 - q^2 \right) - \frac{\epsilon}{3} q^3.
\end{equation}
This represents a simple harmonic oscillator (with natural period
$2\pi$) with an additional force $F(q) = - \epsilon q^2$.  For
$\epsilon \ll 1$, the system is amenable to solution using the methods
from the previous section.  In particular, because the perturbation
term is a polynomial in $q$, we can easily compute the
$\order{\epsilon^2}$ discrete Lagrangian:
\begin{multline}
  L_D\left(h, q_0, q_1\right) = \int_0^h dt\, L\left(q_A\left(t; q_0,
      q_1 \right), D q_A\left(t; q_0, q_1 \right)\right) = \\
  \frac{1}{2}\left( q_0^2 + q_1^2\right) \cot(h) - q_0 q_1 \csc(h) \\
  - \epsilon \left[ \left(q_0 + q_1\right) \left(2 q_0^2 + q_0 q_1 + 2
      q_1^2\right) + \left( q_0^3 + q_1^3 \right) \cos(h) \right]
  \sec^2\left(\frac{h}{2}\right) \tan\left(\frac{h}{2}\right)
\end{multline}

\begin{figure}
  \begin{center}
    \includegraphics[width=0.75\textwidth]{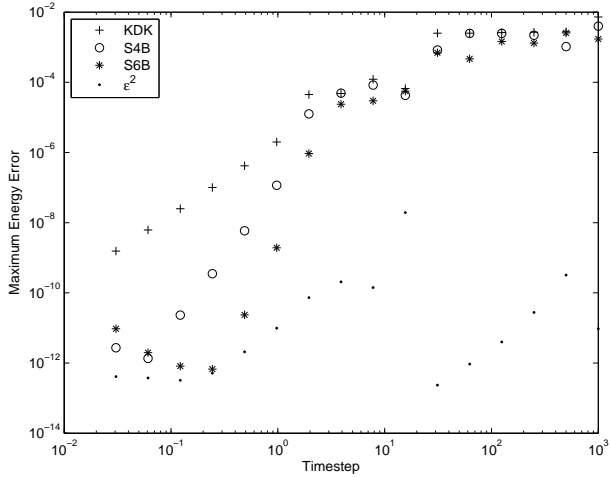}
  \end{center}
  \caption{\label{PerturbedSHOFigure} Maximum relative energy error in
    a simulation of the perturbed SHO over a total time $T = 1000$ as
    a function of timestep, $h$ by the methods of Section
    \ref{IntegratorSection}.  The data are for KDK (plus), S4B
    (circle), S6B (star), and $\order{\epsilon^2}$ variational (dot).
    We set $\epsilon = 10^{-5}$.  Though it is an implicit method, and
    takes more work per step, the $\order{\epsilon^2}$ variational
    integrator has such dramatically better error behavior at large
    stepsize that it is more efficient for integrating this system.}
\end{figure}

Figure \ref{PerturbedSHOFigure} displays the maximum energy error over
a simulation of the oscillator with a total time $T = 1000$ as a
function of the timestep $h$ for the various methods in Section
\ref{IntegratorSection}.  We can see in Figure
\ref{PerturbedSHOFigure} that the $\order{\epsilon^2}$ variational
method significantly outperforms the other methods at large timesteps.

\subsection{Jupiter, Saturn, and the Sun}
\label{JupiterSaturnSunSection}

This subsection reports on simulations of the Jupiter-Saturn-Sun
system with realistic initial conditions.  The Lagrangian for this
system is 
\begin{equation}
  L = \frac{1}{2} \left(m_\odot v_\odot^2 + m_J v_J^2 + m_S v_S^2 
  \right) + \frac{G m_\odot m_J}{r_{\odot J}} + \frac{G m_\odot
    m_S}{r_{\odot S}} + \frac{G m_J m_S}{r_{JS}}.
\end{equation}
The well-known Jacobi transformation (see, e.g.\ \citet{Wisdom1991})
can transform this Lagrangian into a sum of center-of-mass motion, two
Kepler Lagrangians, and perturbing terms with magnitude $\epsilon \sim
m_J/m_\odot \sim 10^{-3}$.  

In this paper, we evaluate the time-average of the perturbing
terms---which are essentially the disturbing function for the
three-body problem---for the $\order{\epsilon^2}$ variational
integrator on the $q_A$ (Kepler) trajectory using numerical
quadrature.  Numerical quadrature is adaptive; each quadrature point
corresponds to a (weighted) kick at that time, and quadrature points
are allocated non-uniformly on the interval $[0,h]$ to best
approximate the integral.  With this technique, we can use the
$\order{\epsilon^2}$ variational method on highly elliptical orbits
with a large timestep without loss of accuracy, since the quadrature
routine will allocate points densely near pericenter.  The other
integration methods, which allocate quadrature points at fixed
fractions of the stepsize, must be run with a small enough stepsize to
resolve rapidly changing forces near pericenter passage throughout the
entire orbit.

\begin{figure}
  \begin{center}
    \includegraphics[width=0.75\textwidth]{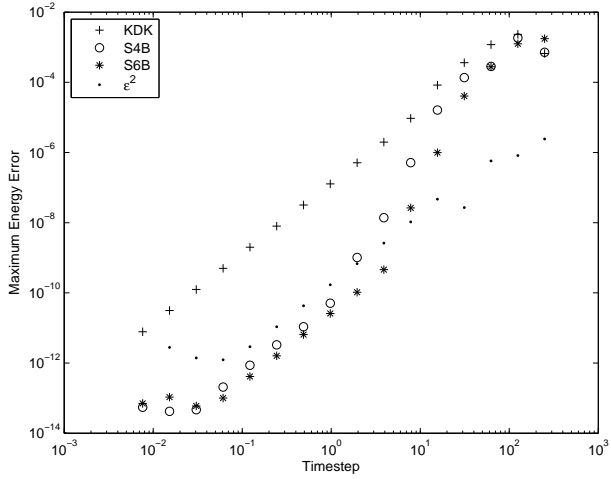}
  \end{center}
  \caption{\label{SolarSystemFigure} Maximum relative energy error
    versus stepsize in a simulation of Jupiter, Saturn and the Sun's
    mutual gravitational interaction over a period of 250 years
    (Jupiter's period is about 12 years).  The curves are for the
    following algorithms: KDK (plus), S4B (circles), S6B (stars), and
    $\order{\epsilon^2}$ variational (dots).  Because in this system
    $\epsilon \sim \frac{m_J}{m_{\odot}} \sim 10^{-3}$ is relatively
    large, the $\order{\epsilon^2}$ variational integrator is only
    advantageous at energy errors of order $\epsilon^2 \sim 10^{-6}$.
    However, for such large energy errors, the $\order{\epsilon^2}$
    variational integrator can take stepsizes which are on the order
    of 10 orbital periods, significantly larger than traditional
    algorithms; these large stepsizes more than offset the increased
    computational cost of the method.}
\end{figure}

Figure \ref{SolarSystemFigure} displays the energy error in a
simulation of the Jupiter-Saturn-Sun system for approximately 20
Jupiter orbits (which corresponds to approximately 240 years) versus
timestep.  For a maximum tolerable error of $\epsilon^2 \sim 10^{-6}$,
the $\order{\epsilon^2}$ variational integrator can take stepsizes of
order 10 orbits, while the other methods do not perform well until
there are several kicks per orbit.  However, for high-accuracy
integrations the errors in the methods (excepting KDK) are comparable,
and the extra cost of the averaging in the $\order{\epsilon^2}$
variational integrator when compared to the other methods makes it
sub-optimal.

\subsection{Small-mass Jupiter, Saturn, and the Sun}
\label{BHSection}

This subsection reports on a simulation with the same initial
conditions as Section \ref{JupiterSaturnSunSection}, but with the
masses of Jupiter and Saturn reduced by a factor of $10^{-3}$.  This
brings $\epsilon \sim 10^{-6}$, roughly in line with the size of the
perturbing interaction one might find in a cluster of stars around a
super-massive black hole in the center of a galaxy.

\begin{figure}
  \begin{center}
    \includegraphics[width=0.75\textwidth]{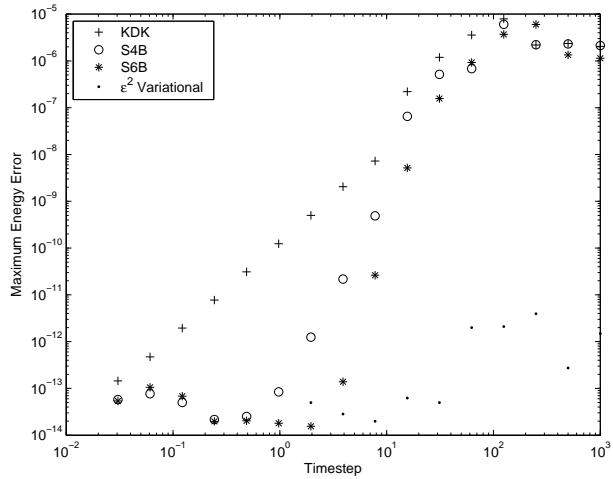}
  \end{center}
  \caption{\label{BlackHoleFigure} Maximum relative energy error in a
    simulation analogous to the one in Section
    \ref{JupiterSaturnSunSection}, except with Jupiter and Saturn's
    masses reduced by a factor of $10^{-3}$ (note that Jupiter still
    has a 12-year period).  The curves are for the following
    algorithms: KDK (plus), S4B (circles), S6B (stars), and
    $\order{\epsilon^2}$ variational (dots).  In this system $\epsilon
    \sim 10^{-6}$, and we see that the $\order{\epsilon^2}$
    variational integrator can take stepsizes which are $\sim 10^3$
    larger than other algorithms for an error budget of $10^{-12}$.}
\end{figure}

Figure \ref{BlackHoleFigure} presents the relative energy error versus
timestep for a simulation of this smaller-$\epsilon$ system over
approximately 100 Jupiter orbits.  In this circumstance, the
$\order{\epsilon^2}$ variational integrator significantly outperforms
the other integrators, even for the (relatively severe) error budget
of $10^{-12}$.  It can take steps which are approximately $10^3$
longer than those of the other integrators at the same energy error
budget, more than compensating for the expensive time-averaging and
implicit nature of the algorithm.

\begin{figure}
  \begin{center}
    \includegraphics[width=0.75\textwidth]{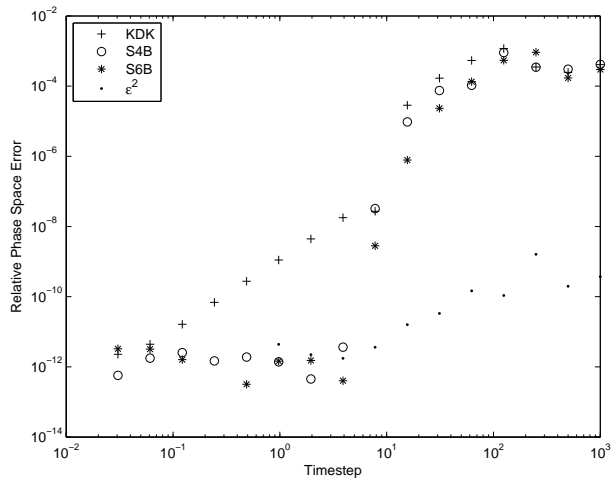}
  \end{center}
  \caption{\label{BlackHoleTrajectoryErrorFigure} Relative phase-space
    (trajectory) error at the end of a simulation analogous to the one
    in Section \ref{JupiterSaturnSunSection}, except with Jupiter and
    Saturn's masses reduced by a factor of $10^{-3}$ (note that
    Jupiter still has a 12-year period).  The curves are for the
    following algorithms: KDK (plus), S4B (circles), S6B (stars), and
    $\order{\epsilon^2}$ variational (dots).  In this system $\epsilon
    \sim 10^{-6}$, and we see that the $\order{\epsilon^2}$
    variational integrator can take stepsizes which are $\sim 10^3$
    larger than other algorithms for a phase-space error budget of
    $10^{-10}$.}
\end{figure}

In Figure \ref{BlackHoleTrajectoryErrorFigure} we plot the trajectory
error of the various methods at the end of the the simulation period
(100 Jupiter orbits).  The $\order{\epsilon^2}$ variational method
outperforms the other methods by approximately a factor of $10^3$ in
stepsize at a relative trajectory error of $10^{-10}$.    

\section{Conclusion}

The variational framework subsumes standard symplectic methods.  In
this work, we have presented the pseudo-high-order integrators of
\citet{McLachlan1995}, \citet{Chambers2000}, and \citet{Laskar2001}
from the variational viewpoint for systems with Lagrangian $L =
L^{(A)} + \epsilon L^{(B)}$.  In addition, we have used the
variational framework to derive a novel implicit integrator which uses
the average perturbing Lagrangian over trajectories of the dominant
Lagrangian to remove all errors from the integration at
$\order{\epsilon}$.  We have presented numerical evidence that, for
small $\epsilon$, this latter integrator is more efficient than
standard pseudo-high-order symplectic integrators for perturbed
systems.  It would be interesting to investigate the performance of
this latter integrator with various analytical approximations to the
average of the perturbing Lagrangian.

\begin{acknowledgements}
  I wish to thank Edmund Bertschinger and Jack Wisdom for helpful
  comments on this work, and Scott Tremaine and Piet Hut for their
  comments and the kind invitation to visit the Institute for Advanced
  Study, where I began this work.  This work was supported by NSF
  grant AST-0407050 and NASA grant NNG06-GG99G.
\end{acknowledgements}

\bibliography{kv}

\end{document}